\documentclass[12pt,preprint]{emulateapj}

\usepackage{graphicx,amsmath,natbib}

\slugcomment{Accepted to the Astrophysical Journal (June 15th, 2009)}

\shorttitle{An image of the HR4796A disk achieved through speckle suppression}
\shortauthors{Hinkley et al.}

\begin{document}

\title{Speckle Suppression Through Dual Imaging Polarimetry, and a Ground-Based Image of the HR 4796A Circumstellar Disk} 

%\title{Ground Based Detection of the faint HR4796A circumstellar disk through dual-channel imaging polarimetry}

\author{Sasha Hinkley\altaffilmark{1,2}}
%\affil{\it Department of Astronomy, Columbia University, 550 West 120th Street, New York, NY  10027, electronic address: shinkley@astro.columbia.edu}

\author{Ben R. Oppenheimer\altaffilmark{2}} 
\author{R\'emi Soummer\altaffilmark{3}} 
\author{Douglas Brenner\altaffilmark{2}}
\author{James R. Graham\altaffilmark{4}}
\author{Marshall D. Perrin\altaffilmark{5}}
\author{Anand Sivaramakrishnan\altaffilmark{2,6,7}}
\author{James P. Lloyd\altaffilmark{8}}
\author{Lewis C. Roberts Jr.\altaffilmark{9}}
\author{Jeffrey Kuhn\altaffilmark{10}}
%\author{James P. Lloyd}
%\affil{\it Astronomy Department, 230 Space Sciences Building, Cornell University, Ithaca, NY 14853}
%\author{Neil Z.?}

%\altaffiltext{1}{Electronic address: shinkley@astro.columbia.edu}
\altaffiltext{1}{Department of Astronomy, Columbia University, 550 West 120th Street, New York, NY  10027, electronic address: shinkley@astro.columbia.edu}
\altaffiltext{2}{Astrophysics Department, American Museum of Natural History, Central Park West at 79th Street, New York, NY 10024}
\altaffiltext{3}{Space Telescope Science Institute, 3700 San Martin Drive, Baltimore, MD 21218}
\altaffiltext{4}{Department of Astronomy, 601 Campbell Hall, University of California Berkeley, CA 94720}
\altaffiltext{5}{NSF Postdoctoral Fellow, UCLA Department of Astronomy}
\altaffiltext{6}{NSF Center for Adaptive Optics.}
\altaffiltext{7}{Stony Brook University}
\altaffiltext{8}{Department of Astronomy, Cornell University, Ithaca, NY 14853}
\altaffiltext{9}{Jet Propulsion Laboratory, California Institute of Technology, 4800 Oak Grove Dr., Pasadena CA 91109}
\altaffiltext{10}{Institute for Astronomy, University of Hawaii, 2680 Woodlawn Drive, Honolulu, Hawaii 96822}

\begin{abstract}
We demonstrate the versatility of a dual imaging polarimeter working in tandem with a Lyot coronagraph and Adaptive Optics to suppress the highly static speckle noise pattern---the greatest hindrance to ground-based direct imaging of planets and disks around nearby stars.  Using a double difference technique with the polarimetric data, we quantify the level of speckle suppression, and hence improved sensitivity, by placing an ensemble of artificial faint companions into real data, with given total brightness and polarization. For highly polarized sources within 0.5$^{\prime\prime}$, we show that we achieve 3 to 4 magnitudes greater sensitivity through polarimetric speckle suppression than simply using a coronagraph coupled to a high-order Adaptive Optics system.  Using such a polarimeter with a classical Lyot coronagraph at the 3.63-m AEOS telescope, we have obtained a 6.5$\sigma$ detection in the $H$-band of the 76 AU diameter circumstellar debris disk around the star HR 4796A. Our data represent the first definitive ground-based near-IR polarimetric image of the HR 4796A debris disk and clearly show the two outer ansae of the disk, evident in Hubble Space Telescope NICMOS/STIS imaging. Comparing our peak linearly polarized flux with the total intensity in the lobes as observed by NICMOS, we derive a lower limit to the fractional linear polarization of $>29\%$ caused by dust grains in the disk. In addition, we fit simple morphological models of optically thin disks to our data allowing us to constrain the dust disk scale height ($ 2.5^{+5.0}_{-1.3}$ AU) and scattering asymmetry parameter ($g$=$<$$\cos\theta$$>$$=0.20^{+.07}_{-.10}$).  These values are consistent with several lines of evidence suggesting that the HR 4796A disk is dominated by a micron-sized dust population, and are indeed typical of disks in transition between those surrounding the Herbig Ae stars to those associated with Vega-like stars. 
\end{abstract}



%% and \citet commands to identify citations. The citations are

\section{Introduction}
Direct imaging surveys of the close-in ($\lesssim$ 1000AU) environments of nearby stars for companions and circumstellar disks are coming into maturity and returning spectacular results  \citep{ldm07,ncb08,kgc08,mmb08,obh08, mgp08, mh09}. Much of the technical effort in high-contrast imaging centers around suppressing the overwhelming luminosity of the host star, and the residual speckle noise, largely caused by uncorrected aberrations in the incoming wave front. This quasi-static source of noise, an especially bad problem for ground-based efforts, is stable for minutes or hours \citep{hos07} and is the largest obstacle to direct detection of companions or circumstellar disks \citep{rwn99,slh02,mdn03,sfa07}. Many authors have suggested useful techniques for direct subtraction of speckle noise through image post processing \citep{sf02,mld06,hos07}.  Another technique using a dual-channel imaging polarimeter can extract a polarized signal due to cirumstellar material (companions or a disk) from the array of unpolarized speckles \citep{kpp01,p03,pgk04,obh08}.  The Lyot Project \citep{osm03,odn04,soh07} employed this technique in addition to using a very high-order Adaptive Optics (AO) system \citep{rn02} and a classical Lyot coronagraph working at the diffraction limit \citep{l39,m96,skm01}.  

In this work, we carry out a sensitivity analysis for speckle suppression through polarimetry, and quantify the level of suppression (Section 2). In Section 3 we demonstrate the power of this technique with a detection of  the debris disk surrounding the nearby ($d$ = 72.8 $\pm$ 1.7 pc), young (8 $\pm$ 2 Myr) star HR 4796A (A0 V, $V$=5.78 mag, \citet{j91,krw98,alm99,ssb99,v07}), achieved in the $H$-band from the ground. To our knowledge, this is the first high-contrast polarimetric near-IR image of the the HR 4796A disk obtained from a ground-based observatory.  This technique is especially well suited for regions of the image which are heavily dominated by speckle noise.  Although circumstellar disk imaging has largely been performed using space-based observatories \citep{ksm00,shc04,kgc05,gkm07}, this paper demonstrates the power of speckle suppression and its ability to image circumstellar disks from ground-based observatories. 

\begin{figure*}[ht]
\center
\resizebox{1.05\hsize}{!}{\includegraphics{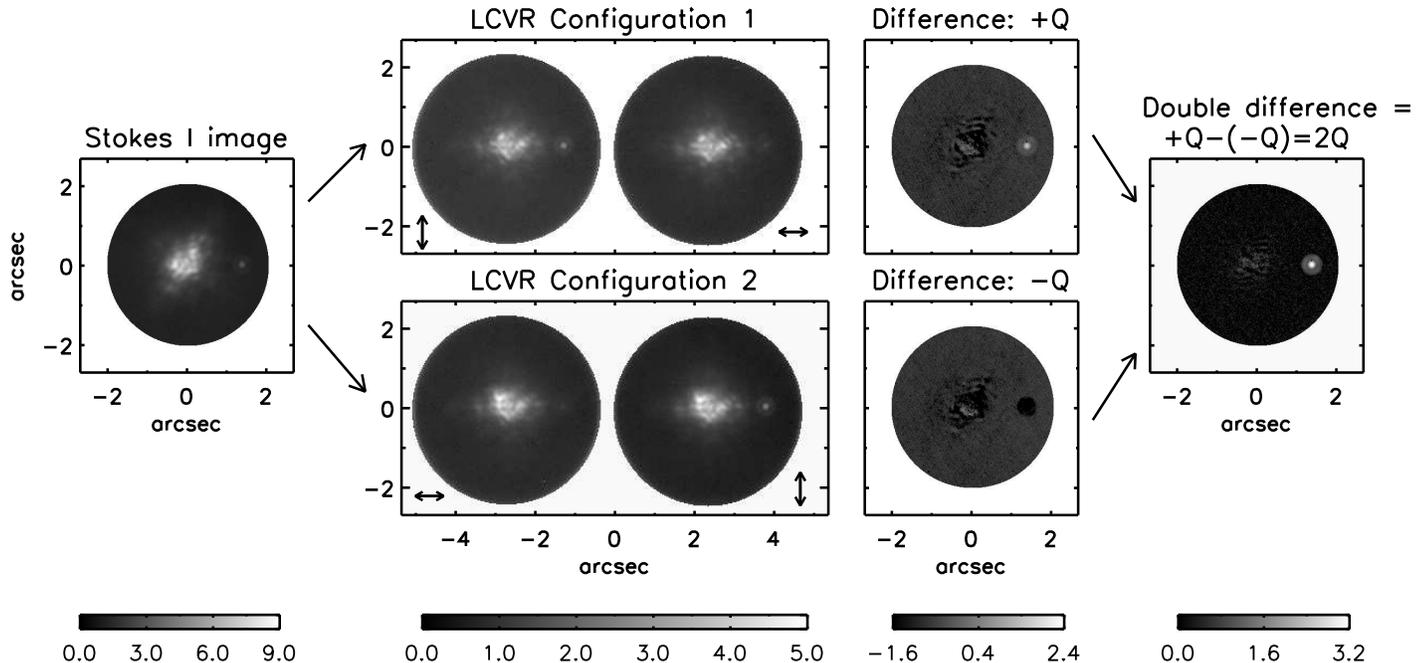}}
  \caption{A schematic illustration of the steps involved in the reduction of data taken with a dual-channel imaging polarimeter. A faint 80\% polarized artificial companion has been placed at the 3 o' clock position.  The ``LCVR Configuration 1'' panel shows the two orthogonal polarization images produced onto the detector by the Wollaston prism, here labelled by the vertical and horizontal arrow symbols.  A difference of the left and right images gives a measure of the Stokes $Q$ amplitude and eliminates the strong speckle halo common to both the left and right images.  Modulating the polarization states (``LCVR Configuration 2'') switches the placement of the two orthogonal images, and the resulting subtraction of these images gives the $-Q$ amplitude.  A difference of the $+Q$ and $-Q$ images eliminates those aberrations not common to the left and right channels, while preserving any polarized signal (see text). The images have a square-root scaling, and the scale bars reflect the increasing sensitivity as the reduction progresses. }
  \label{dbldiff}
\end{figure*}

\section{Speckle Suppression through Polarimetric Observations}
The advantages of differential imaging polarimetry for high-contrast observations of circumstellar disks have previously been discussed at length by several authors \citep{kpp01,p03,hgb06,pgk04,pgl08}. We briefly repeat here some basic concepts to establish consistency and notation. The polarization of light is usually represented by the
Stokes vector $[I, Q, U, V]$\citep{s1852, c46}. The usual astronomical convention is for the $+Q$ direction to be oriented north-south, and $+U$ northeast-southwest, with angles increasing counterclockwise from north to east.  Linear polarization can also be expressed in terms of polarized intensity, $(Q^2 + U^2)^{1/2}$ , and position angle $\theta = 1/ 2\arctan{(U /Q)}$. 
%For astrophysical situations involving the scattering of light, circular polarization is usually (though not always) small compared to linear polarization, so we shall generallydrop Stokes $V$ . 
The normalized polarized intensity $(Q^2 + U^2 + V^2)^{1/2}/I$ is referred to as the degree of polarization, polarization fraction, or percent polarization. Notation is not always consistent: some authors use $P$ to refer to polarized intensity while others use it for degree of polarization. In this work, capital $I$, $Q$, $U$, $V$ , and $P$ will always refer to intensities (e.g., with units of Janskies or Jy arcsec$^{-2}$ ), not normalized quantities.  \citet{t96} and \citet{k02} provide excellent introductions to astronomical polarimetry, while \citet{aad05} summarize the recent state of the art.

At its simplest level, a dual-channel differential imaging polarimeter consists of any device that splits an image into two orthogonal polarization states. This is frequently achieved through the use of a Wollaston prism, a two-element birefringent prism which separates an incoming beam into two orthogonal polarization states, while at the same time introducing an angular deflection between the two beams.  The project described in this paper used such a configuration.  A typical image showing the two displaced fields (left and right) resulting from the beam deflection, and their directions of polarization is shown in Figure~\ref{dbldiff}. A measurement of the Stokes $Q$ parameter can be obtained through a difference of the left and right channels. Such a subtraction (``Difference: +$Q$'' in Figure~\ref{dbldiff}) largely eliminates the unpolarized speckle halo common to both left and right channels. However, to eliminate the bulk of the remaining aberrations (aberrations not common to both channels) that persist in this difference image, we obtain a second measurement by modulating the polarization states by $90^\circ$.  Subtracting these in turn gives a $-Q$ image. We modulate the polarization through the use of Liquid Crystal Variable Retarders (LCVRs). This $-Q$ image is obtained by swapping the positions of the polarization states and subtracting the two channels.  After the subtraction, any astrophysical object will now possess negative counts in the image, but those non-common path aberrations will have the same sign and spatial characteristics present in the $+Q$ image. Subtracting the $-Q$ image from the $+Q$ image (ideally) eliminates the non-common aberrations, leaving only the astrophysically interesting targets present.  Different modulations of the polarization states can be used to obtain Stokes $U$ and $V$.  Since no photons are lost in this process, the Stokes $I$ image can be obtained by summing the left and right images of any polarization configuration.  A schematic representation of the reduction process for a full polarimetric sequence is shown in Figure~\ref{dbldiff}.  

\subsection{Observations}
Under very good observing conditions on 2005 January 26 UTC at the 3.63 m Advanced Electro-Optical System (AEOS) telescope in Maui, we obtained three $H$-band (1.65 $\mu$m) coronagraphic polarimetric sequences ($+Q$, $-Q$, $+U$ modes) of the star HR 4796A.  The data were gathered using the Lyot Project coronagraph \citep{osm03,odn04,soh07}  and the Kermit infrared camera \citep{p02}. The coronagraph was a diffraction-limited,  classical Lyot coronagraph \citep{l39,m96} with a 455 mas diameter occulting mask and a hard-edged Lyot stop.  The AEOS telescope is an altitude-azimuth design, equipped with an AO system using a  941 actuator deformable mirror \cite{rn02}.  The total observing time for this dataset was 1080 s, comparable to the 1024 s for the \citet{ssb99} F160W HST data. During the observations, the local Fried parameter, $r_0$, a measure of the strength of turbulence in the atmosphere above the observatory, spanned the range of 15 to 25 cm, indicating nearly ideal conditions for AO observations at AEOS.  Over the course of the observations, we obtained only +$Q$, -$Q$, and +$U$ images because our retarders did not provide sufficient retardance to obtain a $-U$ image. 

\begin{deluxetable}{ccccc}
\tabletypesize{\scriptsize}
\tablecaption{Table of HR 4796A Observations on UTC 2005 January 26}
\tablewidth{0pt}
\tablehead{\colhead{Polarization Mode}  & \colhead{Filter} &\colhead{Exp. Time}  & \colhead{No. of Images} &\colhead{Total}}
\startdata                                     
  +$Q$    & $H$ & 120 s  & 3 & 360 s                  \\
   -$Q$    & $H$ & 120 s  & 3 & 360 s                  \\
   +$U$    & $H$ & 120 s  & 3 & 360 s                  \\
    Total   &          &             &    & 1080 s
\enddata
\label{obstable}
\end{deluxetable}

The polarimeter implemented in the Lyot Project for the data in this paper was unique in two regards.  First, the Wollaston prism was located immediately after the Lyot pupil in the coronagraph.  This post-pupil location is the correct location to minimize differential aberrations between the two beams.  This setup is an improvement over other polarimeters designed for use in high-contrast imaging, e.g. \citet{pgl08}.  Second, the use of LCVRs as a polarization modulator is relatively rare for night-time polarimetry.  The benefits of using LCVRs include a great deal of increased flexibility in modulation, a lack of any image motions induced by rotating optics, and slightly faster modulation (although still not faster than the atmospheric timescales involved).  Disadvantages of using retarders of this type include a potentially reduced wavefront quality.  
%In addition, during the particular observing sequence of HR 4796A described in the next section, we were unable to obtain the $-U$ modulation state due to insufficient retardance, although other datasets used all modes \citet{obh08}.  

\subsection{Polarimetric Dynamic Range}
\begin{figure}[ht]
\center
\resizebox{1.00\hsize}{!}{\includegraphics{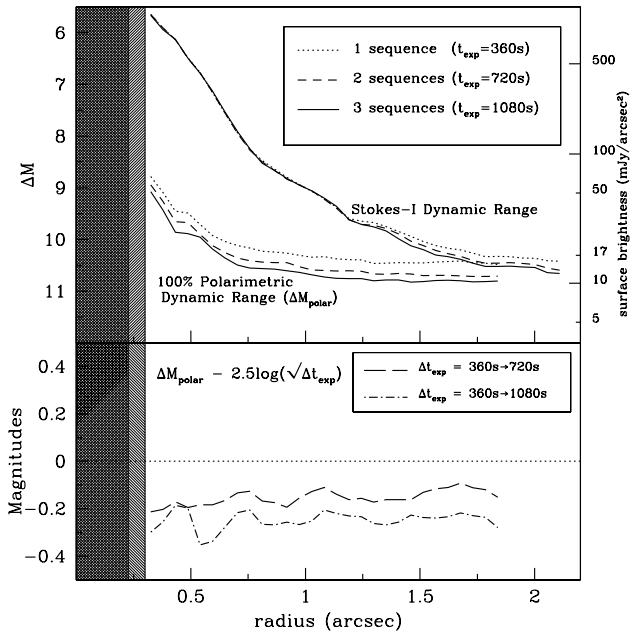}}
  \caption{
  Top panel: The three lines labelled ``Stokes-$I$ Dynamic Range'' show the faintest detectable $H$-band $5\sigma$ point source, relative to the host star, as described in \citet{hos07} and \citet{sfa07}. Each line indicates a co-addition of data with equivalent exposure time of 360 s, 720 s, and 1080 s. This sensitivity is limited by the bright, uncorrected speckle halo in the field-of-view.  The collection of three lines labelled ``100\% Polarimetric Dynamic Range ($\Delta M_{\hbox{polar}}$)'' show the faintest detectable 100\% polarized source using our dual-imaging polarimeter. The dual-imaging polarimeter removes a majority of the unpolarized speckle halo, and the polarimetric sensitivity increases with exposure time at a rate close to that expected for uncorrelated Gaussian noise.  The lower panel shows the deviation of this improved polarimetric sensitivity, through co-adding two and three total polarimetric sequences, from the expected $\sqrt{\Delta t_{\hbox{exp}}}$ gain attributed to purely Gaussian noise. The inner grey region shows the size of our occulting mask, while the wider grey region shows where the suppression of the speckle noise was not effective.  } 
  \label{polardr}
\end{figure}

Our goal is to quantify the gain in dynamic range achievable using the dual-imaging
polarimetry technique. As a reference, we start by illustrating the dynamic range achieved on our Stokes $I$ images without taking advantage of the polarimetric capabilities.  According to a technique we discussed previously \citep{sfa07,hos07}, we have derived the magnitude difference (dynamic range) between the occulted star and a $5\sigma$ point source as a function of position in the Stokes $I$ images. The residual scattered light outside of our coronagraphic mask, usually in the form of highly persistent speckle noise, is the main limiting factor for detection of a point source (See Section 2.3.1). Local evaluation of the amplitude of this  noise in turn determines the minimum brightness required for a 5$\sigma$ point source detection.  A radial curve of this sensitivity is shown in Figure~\ref{polardr} and labelled ``Stokes-I Dynamic Range'.' 

%To determine our sensitivity with the double difference technique (and also our sensitivity with the HR 4796A observation), we chose for our test data a ``control'' observation of the star HD 107146 ($V$=7.07 mag, G2 V).  We chose to use a dataset other than HR 4796A to avoid the actual disk structure biasing our estimates of the sensitivity.  These data were taken in a manner identical to that of HR 4796A on 2006 December 12 UTC.  There were no changes to the system between these observations, the observing conditions were equivalently good, and the AO system provided similar quality of correction on both targets.  This star was a member of our original target list, due to its known debris disk.  However, given that the disk is optically very thin, and has only been imaged outside of the field of view of our observations \citep{agk04}, no disk structure was observed in our field of view.  We are confident that our sensitivity estimations using this star can be applied to the HR 4796A dataset.  A polarimetric image of HD 107146 is shown in \citet{obh08}.  In addition, although eight sequences identical to the one depicted in Figure~\ref{dbldiff} were obtained for HD 107146, our sensitivity estimations here use only three coadded sequences, so as to replicate the three coadded sequences taken for HR 4796A (see Section 3).   

All photometric values in this paper were calibrated to unocculted images of HR 4796A, directly prior to the occulted sequences. The raw data images were calibrated through standard dark current subtraction, application of bad pixel maps,  and flat-fielding.  The flat field images were acquired using incandescent lamps each night.   Also, binary star observations with well known orbits were observed to calibrate the pixel scale and image rotation fiducials.  Prior to the subtraction, the images are rotated so that north is up in the image, east is to the left, and registered to each other using a cross correlation with subpixel accuracy.  The data reduction technique is discussed in greater detail in \citet{soh06} and Appendix A of \citet{obh08}

To actually calculate the polarimetric sensitivity on the image, artificial point sources with varying total brightness and degree of polarization were placed radially into the mutually orthogonal left and right images in the original data. An example of this is shown in Figure~\ref{dbldiff}, which shows an artificial 80\% polarized (in the vertical direction) faint point source with a given total brightness. The relative flux for the artificially placed companion in the left and right channels is given by:
 
\begin{eqnarray}
 F_{\updownarrow} &=& \frac{F_{total}}{2}(1 - p\cos{2\theta}) \\
 F_{\leftrightarrow} &=& \frac{F_{total}}{2}(1 + p\cos{2\theta}),  
\end{eqnarray}

where $F_{total}$ is the total flux of the object ($F_{total} = F_{\updownarrow}  +  F_{\leftrightarrow} $).  These images were processed through the normal double differencing process schematically shown in Figure~\ref{dbldiff}, and the total flux of these point sources in the original images were incrementally increased from zero until a $5\sigma$ detection was obtained in the double difference images.  This threshold {\it total} intensity was recorded and is plotted in Figure~\ref{polardr} labelled ``100\% Polarimetric Dynamic Range ($\Delta M_{\hbox{polar}}$).''

\subsection{Results}
Figure~\ref{polardr} shows the speckle suppression achieved with the double differencing technique is most effective within $1.75^{\prime\prime}$ and can achieve up to 4 magnitudes of improvement over the Stokes $I$ within $0.5^{\prime\prime}$.  This level of polarimetric suppression is comparable to other AO-based polarimeters \citep{pcr00,pgk04,pgl08}, but results using this particular polarimeter on AEOS benefit from the extremely high-order AO system \citep{rn02}, and the optimized coronagraph \citep{skm01} in the beam path before any of the polarimetry optics. 

Modelling point sources are especially useful in the context of the data discussed in the next section, since the HR 4796A disk ansae are very near to point sources. Finding the required polarization for detection in our double-difference technique will thus help us to further constrain the value obtained directly (using published NICMOS Stokes $I$ values) described in Section 3.  
These calculations using point sources can be directly applied to extended objects with resolved surface brightnesses.  We have recasted the sensitivity results in terms of surface brightness (mJy/arcsec$^2$), and those values are listed on the right-hand axes of Figure~\ref{polardr}.  It should be noted, though, that in Figure~\ref{polardr} the brightness difference values (left-hand axis) indicate the total brightness of  a point source, while the surface brightness (mJy/arcsec$^2$) values reflect the brightness of the peak intensity of the point source. Nonetheless, our analysis for point sources translates over to extended sources since the key issue is the overall brightness of the source in comparison to the amplitude of the surrounding noise.  

\subsubsection{Post Speckle Suppression: A Return to the Read Noise Limited Regime}
The uncorrected speckle noise in high contrast imaging data is due to highly static aberrations in the incoming wave front that are not corrected by the AO system, or are induced ``downstream'' from the wave front sensor.  Consequently, the resulting speckle noise in the images is highly stable with time. As \citet{hos07} demonstrate (esp. their Fig. 2), since the speckle noise pattern persists with sequential images, simple co-adding of data does not significantly improve the coronagraphic sensitivity, as would be expected for uncorrelated Gaussian-type noise. The speckles must be removed to gain improvements in sensitivity. \citet{mld06} have performed speckle subtraction through image post-processing to greatly enhance their dynamic range, while \citet{hob08} employ an instrument which uses the wavelength dependence of the speckles to disentangle them from a true astrophysical source. In this work, we use a dual-imaging polarimeter to subtract the unpolarized speckle pattern from the images. 

Once the highly-static, highly time-correlated, speckle noise has been removed through polarimetry, the resulting noise characteristics in the double difference image are distinctly similar to noise with Gaussian type properties.  This is reminiscent of the read noise dominated regime in which speckles are not the dominant source of noise. In this regime, normal Gaussian-like noise properties of  the image become applicable, and sensitivity to polarized sources should increase with the square root of exposure time.  Indeed, our data show behavior quite close to this.  The dashed and solid lines of the polarimetric dynamic range curves shown in Figure~\ref{polardr} show the sensitivity with double and triple the effective exposure time of that represented by the dotted line. If the image noise that dictates the sensitivity is similar to Gaussian type noise, the expected gain in sensitivity is $2.5\log{\sqrt{2}}\simeq0.38$ mag and $2.5\log{\sqrt{3}}\simeq0.60$ mag. In the lower panel, we show the deviation in the curves from this expected gain. These residual curves are within 0.1 - 0.2 magnitudes of zero, consistent with a noise pattern with largely Gaussian properties. Moreover, the fact the the residual is consistently negative indicates that the speckle noise pattern is nearly, but not quite in the Gaussian regime.  This is due to any residual speckle pattern that was not completely subtracted during the double difference process.

\section{Detection of the HR4796A debris disk}
\begin{figure}[ht]
\center
\resizebox{1.1\hsize}{!}{\includegraphics{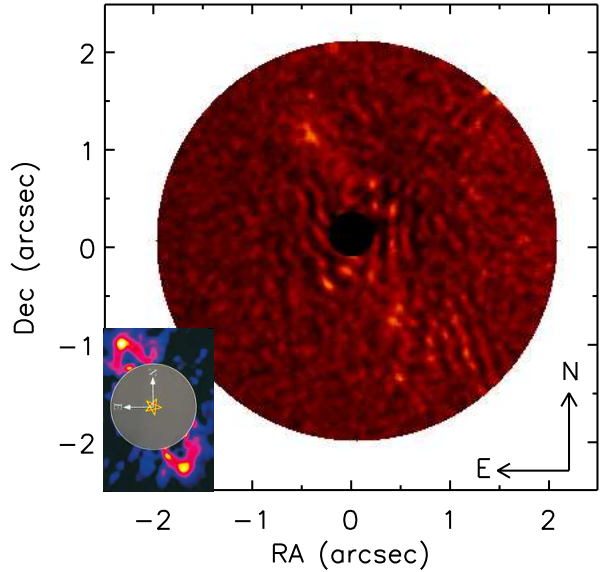}}
  \caption{An $H$-band fractional polarization ($P_{\hbox{linear}}=\sqrt{Q^2 + U^2}/I$) image for the star HR 4796A taken using the Lyot Project dual-imaging polarimeter at the AEOS telescope. The extreme lobes (ansae) of the circumstellar debris disk are detected at the 6.5$\sigma$ level and show a lower limit fractional polarization of $44 \%$ (see text). The central black circle indicates the size of our coronagraphic mask, and the image has been smoothed with a 3 pixel (40 mas) smoothing function.  The image in the lower left corner is an HST $1.6\mu m$ (F160W) image taken from \citet{ssb99}.  }
  \label{hr4796a} 
\end{figure}

Using the method described in the previous section, we have obtained a modest, yet significant detection of the circumstellar disk around HR 4796A  shown in Figure~\ref{hr4796a}.  Although the full ringlike structure of the disk is not immediately evident in our image, we clearly detect the two extreme edges of the disk (ansae) at the 6.5$\sigma$ level above the residual image noise. We measure a position angle of $27.5^\circ \pm 2.5^\circ$, in good agreement with the $27^\circ.01 \pm 0^\circ .16$ as measured by \citet{swb09}.  The intensity in linearly polarized light (shown in figure~\ref{hr4796a}) was determined from our double-difference $Q$ images (described schematically in Figure~\ref{dbldiff}) as well as the trio of Stokes $U$ images to construct a normalized Stokes $P$ image: $P_{\hbox{linear}}=\sqrt{Q^2 + U^2}/I$. We find a peak polarized flux density on the brighter (north) lobe of 7.4 mJy/arcsec$^2$. Comparing this to the published peak brightness (Stokes $I$) from \citet{ssb99} of 17 mJy/arcsec$^2$, we derive a fractional polarization of $ 44 \pm 5\%$, comparable to values found in other debris disks, e.g. AU Mic \citep{gkm07}.  Expressing this as the $3\sigma$ lower limit, we state that the true polarization is greater than 29\%.  However, it should be noted these values are only lower limits to the true fractional linear polarization: recent Lyot Project data from the AEOS telescope suggest some Stokes $V$ polarization induced by the AEOS telescope \citep{obh08,hk08} may be present, reducing the amount of linearly polarized flux in the $Q$ and $U$ modes. Since this particular observing sequence did not have the capabilities to measure Stokes $V$, we present our result merely as a lower limit. Given these issues, along with the relatively small telescope aperture, this is still a significant demonstration of the speckle removal technique.

\begin{deluxetable}{lll}
\tabletypesize{\scriptsize}
\tablecaption{HR 4796A $H$-band Disk Morphology}
\tablewidth{0pt}
\tablehead{\colhead{Study}  & \colhead{Ansae Separation} &\colhead{PA (deg E of N)} }
\startdata                                     
    \citet{ssb99} &  $2.12\pm 0.04^{\prime\prime}$   &  $26.8\pm0.6^\circ$                        \\
        \citet{swb09} &  $2.107\pm 0.0045^{\prime\prime}$   &  $27.01\pm0.16^\circ$        \\
     This work     &  $2.10 \pm 0.05^{\prime\prime}$  &  $27.5\pm 2.5^\circ$      
\enddata
\label{disktable}
\end{deluxetable}

Shown in Figure~\ref{polardr} is the peak surface brightness (17mJy/arcsec$^2$) of the brighter of the two disk ansae from \citet{ssb99}. Inspection of the plot suggests this corresponds to a polarization level of 70\%, still consistent with our direct lower limit calculation of 44\%.  However, such a comparison may not be completely valid, as the dynamic range analysis assumed point sources for the sensitivity derivation while the disk ansae are more extended lobes. 
%Trying to reconcile these two numbers  assumes that the ansae are point sources. In fact, the ansae have a $\sim$70 mas Full Width at Half Maximum (FWHM), as compared to the $\sim90$ mas FWHM for a typical $H$-band point spread function at AEOS.  This difference in morphology can account for the difference in fractional polarization from the artificial point-source method ($\sim90$\%) and the direct lower limit method (44\%). 

\subsection{Morphological Models}
\begin{figure}[ht]
\center
\resizebox{0.70\hsize}{!}{\includegraphics{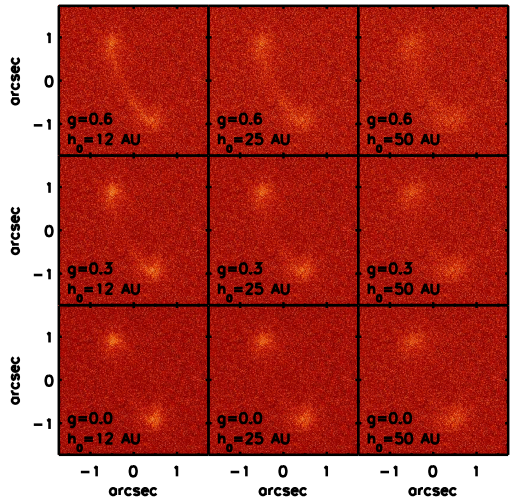}}
  \caption{A matrix of fractional polarization images (same as Figure~\ref{hr4796a}) of our disk models for various morphologies. Each grid column represents a different disk scale height (12, 25, or 50AU); each row shows a different value of the Henyey-Greenstein $g$=$<$$\cos\theta$$>$ parameter (0.0,0.3, and 0.6). The disk position angle, inclination, inner/outer radii were taken from \citet{swb09}. Gaussian noise has been added to match the data in Figure~\ref{hr4796a}. 
%A $\chi^2$ fitting to these models is shown in Figure~\ref{chicontour}. 
}
  \label{modelgrid}
\end{figure}

To constrain the nature of the debris disk around HR 4796A, we model optically thin disks assuming a Henyey-Greenstein phase function \citep{hg41} and a Raleigh-like variation of polarization with scattering angle, which is suitable for disks with small grains or larger grain aggregrates \citep{kkm06}. The model is a two-dimensional generalization of the of the one-dimensional model used in \citet{gkm07}, appropriate for a solar system zodiacal Henyey-Greenstein dust model \citep{h85,gkm07}. We have chosen to fit for the Henyey-Greenstein scattering asymmetry paramter $g$=$<$$\cos\theta$$>$ and the disk scale height.  We fit for these two parameters, since both are intrinsically related to the dust structure in the disk, and can most readily be constrained by the polarimetric data.  A sample grid of models with scale heights $h_0$ = 12AU, 25AU, and 50AU as well as Henyey-Greenstein parameters $g$=0.0, 0.3, and 0.6 is shown in Figure~\ref{modelgrid} to guide the eye.  A value $g=0$ is completely isotropic scattering, while $g=0.6$ signifies moderately strong forward scattering.  A value of $g=0.3$ is typical of Zodiacal dust and some debris disks.  The model assumes a Gaussian vertical density distribution with an adjustable scale height based on the COBE model of zodiacal background light \citep{kwf98}.  Using cylindrical coordinates, we adopt the following density structure for the disk, 

\begin{equation}
  n(r,z) = n_0 \left( \frac{r}{r_0} \right)^{\alpha}e^{-\beta(z/h_0)^\gamma}, 
\end{equation}

with $\alpha = -1.803$, $\beta = 4.973$, $\gamma = 1.265$ motivated by \citet{kwf98}.  We have used the inner and outer radii (69 AU and 83 AU, respectively), inclination (14.12$^\circ$ from edge on), and position angle (63.2$^\circ$) from \citet{swb09}.  Each thumbnail image in Figure~\ref{modelgrid} was computed using the measured pixel scale of the Lyot Project's infrared camera (13.5 mas/pixel) and has been convolved with a PSF for the AEOS telescope at 1.65 $\mu$m.  This PSF also reflects the reduced effective pupil diameter imposed by the Lyot mask in the Lyot Project coronagraph.  During the fits, we also mask out the region covered by our coronagraphic mask.  We generated an ensemble of models varying $h_0$ between 2.5 and 25AU and $g$ between 0.0 and 0.7. 

\subsection{model fit results}
The $\chi^2$ fitting procedure was performed on a pixel-by-pixel basis, normalized with the peak brightness in the model matched to the peak brightness in the polarization image.  Ideally, such a fit should simultaneously use the polarized intensity {\it and } the total intensity of the disk image \citep{gkm07} in the fits to avoid degeneracies between dust properties and disk structure \citep{dmg04,pfm07}.  Given the relatively low signal-to-noise of this detection, and the lack of a detection in a total intensity image, we chose only to perform a $\chi^2$ fit to the polarization image.  Also, the data showing the fractional polarization may give a better handle on the disk scale height.  In addition, the speckle noise pattern in highly corrected AO images can be well modelled with a Rician Probability Distribution Function \citep{as04,fg06,sfa07}, so a model fit should take this fact into account.  However, given that the speckle noise pattern has been significantly subtracted away, as discussed above, a $\chi^2$ minimization employing a Gaussian noise variance is suitable.  

%\begin{figure}[ht]
%\center
%\resizebox{1.30\hsize}{!}{\includegraphics{chitest.ps}}
%  \caption{The $\chi^2$ surface for our model fits to the data. The best fit model is a disk with scale height near 2.5AU, and a Henyey-Greenstein parameter of 0.2. Also shown is the $\chi^2 = 1$ contour. }
%  \label{chicontour}
%\end{figure}

Our best $\chi^2$ fit ($\chi^2\simeq 1.1$) is a model with dust disk scale height of $ 2.5^{+5.0}_{-1.3}$ AU and Henyey-Greenstein phase function of $0.20^{+.07}_{-.10}$. 
%as shown in Figure~\ref{chicontour}.  
These findings are consistent with several lines of evidence suggesting that HR 4796A is dominated by a micron-sized dust population. Such a value of the asymmetry parameter is similar to models employing ISM-type dust grain distributions \citep{wlb02, kmh94}, with the dominant grain size lying between 0.1 and 1 $\mu$m.   This population is similar that of the (8$\pm$2 Myr, A1 V) 49 Cet disk as discussed by \citet{wks07}, with small (0.1 $\mu$m) grains between 30 and 60 AU from the star.  On the other extreme, in their analysis of HR 4796A, \citet{dws08} suggest that the minimum grain sizes for this disk are between 1 and 5 $\mu$m in size, with a dominant population of 1.4 $\mu$m grains.  Although the \citet{dws08} study employed the use of Tholins in their model fits, \citet{kml08}  find that the disk spectra can be well fit using less exotic compounds (amorphous silicates, amorphous carbonates, and water ice).  In these models, 1 $\mu$m grains give the best overall fit.  Another powerful diagnostic of the grain size distribution is the level of linear polarization measured in this work ($44\pm 5\%$).  Given the efficiency with which small grains can polarize starlight, our results are consistent with a micron-sized dust population leading to this relatively high level of polarization.  The age and grain size distribution for HR 4796A support the interpretation of  \citet{wks07}, that it as well as the 49 Cet disk are representative of a class of disks that are ``transitional'' between the Herbig Ae class and Vega-like stars. 

Our best estimate and uncertainties on the disk scale height do not significantly constrain it below 12.5 AU at the 2$\sigma$ level. Although several authors have measured gas scale heights in disks such as $\beta$ Pic \citep{blo04}, to our knowledge this work is one of the only measurements of a dust disk scale height.  \citet{kww99} have simulated the structure of the HR 4796A disk, and predict a scale height between 0.3 and 0.6 AU, nearly consistent with our results.  Our value for the phase function is similiar to the phase function values obtained for other disks, including Fomalhaut \citep{kgc05} and HD 141569A \citep{cka03}.

\section{Conclusions}
In this work, we have demonstrated and quantified the speckle suppression achievable with a dual-channel imaging polarimeter working in conjunction with a high-order AO system and a classical Lyot coronagraph.   Using a double subtraction technique in the data processing, we have demonstrated that a dual-channel imaging polarimeter can facilitate detection of polarized point sources within $\sim$$1^{\prime\prime}$ that are $10 - 100$ times fainter than the faintest (unpolarized) objects detectable by a coronagraph+AO system alone. These results should be of interest for future polarimetry-based planet finding projects \citep{gst04,bfd06,mgp06,lhb09}.  With this instrument, we report the first ground-based near-IR polarimetric detection of the disk surrounding the young star HR 4796A.  Using Stokes $I$ values from NICMOS \citep{ssb99} to complement our data, we derive a lower limit to the fractional linear polarization of $> 29\%$ caused by dust grains in the disk. In addition, we fit simple morphological models of optically thin disks to our data allowing us to constrain the disk dust scale height ($ 2.5^{+5.0}_{-1.3}$ AU) and scattering asymmetry parameter ($g$=$<$$\cos\theta$$>$$=0.20^{+.07}_{-.10}$).  Given these values, and the relatively high level of polarization, these findings are consistent with several lines of evidence suggesting that the HR 4796A disk is dominated by a micron-sized dust population.

%% Included in this acknowledgments section are examples of the
%% AASTeX hypertext markup commands. Use \url without the optional [HREF]
%% argument when you want to print the url directly in the text. Otherwise,
%% use either \url or \anchor, with the HREF as the first argument and the
%% text to be printed in the second.

\acknowledgments
%%{\bf Acknowledgements}

The Lyot Project is based upon work supported by the National Science  
Foundation under Grant Nos. 0334916, 0215793, and 0520822, as well as  
grant NNG05GJ86G from the National Aeronautics and Space  
Administration under the Terrestrial Planet Finder Foundation Science  
Program.  The Lyot Project grateful acknowledges the support of the  
US Air Force and NSF in creating the special Advanced Technologies  
and Instrumentation opportunity that provides access to the AEOS  
telescope.  Eighty percent of the funds for that program are provided  
by the US Air Force.  This work is based on observations made at the  
Maui Space Surveillance System, operated by Detachment 15 of the U.S.  
Air Force Research Laboratory Directed Energy Directorate.
JRG, AS, and MDP were supported in part by the National Science Foundation Science and Technology Center for Adaptive Optics, managed by the University of California at Santa Cruz under cooperative agreement No. AST - 9876783. 
A portion of the research in this paper was carried out at the Jet Propulsion Laboratory, California Institute of Technology, under a contract with the National Aeronautics and Space Administration.
The Lyot Project is also grateful to the Cordelia Corporation, Hilary  
and Ethel Lipsitz, the Vincent Astor Fund, Judy Vale, Alison Cooney, and an anonymous  
donor, who initiated the project. 
%R.S and M.P. are supported by the 
%NASA Michelson Fellowship Postdoctoral/Graduate Fellowship under contract 
%to the Jet Propulsion Laboratory (JPL) funded by BASA. The JPL is managed for 
%NASA by the California Institute of Technology.

%\appendix 

%% curly braces. If you forget this, LaTeX will generate the error
%% different from previous examples. The natbib system solves a host

\bibliography{/Users/shinkley/Desktop/papers/MasterBiblio_Sasha} 

\begin{thebibliography}{66}
\expandafter\ifx\csname natexlab\endcsname\relax\def\natexlab#1{#1}\fi

\bibitem[{{Adamson} {et~al.}(2005){Adamson}, {Aspin}, {Davis}, \&
  {Fujiyoshi}}]{aad05}
{Adamson}, A., {Aspin}, C., {Davis}, C., \& {Fujiyoshi}, T., eds. 2005,
  Astronomical Society of the Pacific Conference Series, Vol. 343,
  {Astronomical Polarimetry: Current Status and Future Directions}, ed.
  A.~{Adamson}, C.~{Aspin}, C.~{Davis}, \& T.~{Fujiyoshi}

\bibitem[{{Aime} \& {Soummer}(2004)}]{as04}
{Aime}, C., \& {Soummer}, R. 2004, \apjl, 612, L85

\bibitem[{{Augereau} {et~al.}(1999){Augereau}, {Lagrange}, {Mouillet},
  {Papaloizou}, \& {Grorod}}]{alm99}
{Augereau}, J.~C., {Lagrange}, A.~M., {Mouillet}, D., {Papaloizou}, J.~C.~B.,
  \& {Grorod}, P.~A. 1999, \aap, 348, 557

\bibitem[{{Beuzit} {et~al.}(2006){Beuzit}, {Feldt}, {Dohlen}, {Mouillet},
  {Puget}, {Antichi}, {Baruffolo}, {Baudoz}, {Berton}, {Boccaletti},
  {Carbillet}, {Charton}, {Claudi}, {Downing}, {Feautrier}, {Fedrigo}, {Fusco},
  {Gratton}, {Hubin}, {Kasper}, {Langlois}, {Moutou}, {Mugnier}, {Pragt},
  {Rabou}, {Saisse}, {Schmid}, {Stadler}, {Turrato}, {Udry}, {Waters}, \&
  {Wildi}}]{bfd06}
{Beuzit}, J.-L., {Feldt}, M., {Dohlen}, K., {Mouillet}, D., {Puget}, P.,
  {Antichi}, J., {Baruffolo}, A., {Baudoz}, P., {Berton}, A., {Boccaletti}, A.,
  {Carbillet}, M., {Charton}, J., {Claudi}, R., {Downing}, M., {Feautrier}, P.,
  {Fedrigo}, E., {Fusco}, T., {Gratton}, R., {Hubin}, N., {Kasper}, M.,
  {Langlois}, M., {Moutou}, C., {Mugnier}, L., {Pragt}, J., {Rabou}, P.,
  {Saisse}, M., {Schmid}, H.~M., {Stadler}, E., {Turrato}, M., {Udry}, S.,
  {Waters}, R., \& {Wildi}, F. 2006, The Messenger, 125, 29

\bibitem[{{Brandeker} {et~al.}(2004){Brandeker}, {Liseau}, {Olofsson}, \&
  {Fridlund}}]{blo04}
{Brandeker}, A., {Liseau}, R., {Olofsson}, G., \& {Fridlund}, M. 2004, \aap,
  413, 681

\bibitem[{{Chandrasekhar}(1946)}]{c46}
{Chandrasekhar}, S. 1946, \apj, 104, 110

\bibitem[{{Clampin} {et~al.}(2003){Clampin}, {Krist}, {Ardila}, {Golimowski},
  {Hartig}, {Ford}, {Illingworth}, {Bartko}, {Ben{\'{\i}}tez}, {Blakeslee},
  {Bouwens}, {Broadhurst}, {Brown}, {Burrows}, {Cheng}, {Cross}, {Feldman},
  {Franx}, {Gronwall}, {Infante}, {Kimble}, {Lesser}, {Martel}, {Menanteau},
  {Meurer}, {Miley}, {Postman}, {Rosati}, {Sirianni}, {Sparks}, {Tran},
  {Tsvetanov}, {White}, \& {Zheng}}]{cka03}
{Clampin}, M., {Krist}, J.~E., {Ardila}, D.~R., {Golimowski}, D.~A., {Hartig},
  G.~F., {Ford}, H.~C., {Illingworth}, G.~D., {Bartko}, F., {Ben{\'{\i}}tez},
  N., {Blakeslee}, J.~P., {Bouwens}, R.~J., {Broadhurst}, T.~J., {Brown},
  R.~A., {Burrows}, C.~J., {Cheng}, E.~S., {Cross}, N.~J.~G., {Feldman}, P.~D.,
  {Franx}, M., {Gronwall}, C., {Infante}, L., {Kimble}, R.~A., {Lesser}, M.~P.,
  {Martel}, A.~R., {Menanteau}, F., {Meurer}, G.~R., {Miley}, G.~K., {Postman},
  M., {Rosati}, P., {Sirianni}, M., {Sparks}, W.~B., {Tran}, H.~D.,
  {Tsvetanov}, Z.~I., {White}, R.~L., \& {Zheng}, W. 2003, \aj, 126, 385

\bibitem[{{Debes} {et~al.}(2008){Debes}, {Weinberger}, \& {Schneider}}]{dws08}
{Debes}, J.~H., {Weinberger}, A.~J., \& {Schneider}, G. 2008, \apjl, 673, L191

\bibitem[{{Duch{\^e}ne} {et~al.}(2004){Duch{\^e}ne}, {McCabe}, {Ghez}, \&
  {Macintosh}}]{dmg04}
{Duch{\^e}ne}, G., {McCabe}, C., {Ghez}, A.~M., \& {Macintosh}, B.~A. 2004,
  \apj, 606, 969

\bibitem[{{Fitzgerald} \& {Graham}(2006)}]{fg06}
{Fitzgerald}, M.~P., \& {Graham}, J.~R. 2006, \apj, 637, 541

\bibitem[{{Gisler} {et~al.}(2004){Gisler}, {Schmid}, {Thalmann}, {Povel},
  {Stenflo}, {Joos}, {Feldt}, {Lenzen}, {Tinbergen}, {Gratton}, {Stuik},
  {Stam}, {Brandner}, {Hippler}, {Turatto}, {Neuhauser}, {Dominik}, {Hatzes},
  {Henning}, {Lima}, {Quirrenbach}, {Waters}, {Wuchterl}, \&
  {Zinnecker}}]{gst04}
{Gisler}, D., {Schmid}, H.~M., {Thalmann}, C., {Povel}, H.~P., {Stenflo},
  J.~O., {Joos}, F., {Feldt}, M., {Lenzen}, R., {Tinbergen}, J., {Gratton}, R.,
  {Stuik}, R., {Stam}, D.~M., {Brandner}, W., {Hippler}, S., {Turatto}, M.,
  {Neuhauser}, R., {Dominik}, C., {Hatzes}, A., {Henning}, T., {Lima}, J.,
  {Quirrenbach}, A., {Waters}, L.~B.~F.~M., {Wuchterl}, G., \& {Zinnecker}, H.
  2004, in Presented at the Society of Photo-Optical Instrumentation Engineers
  (SPIE) Conference, Vol. 5492, Ground-based Instrumentation for Astronomy.
  Edited by Alan F. M. Moorwood and Iye Masanori. Proceedings of the SPIE,
  Volume 5492, pp. 463-474 (2004)., ed. A.~F.~M. {Moorwood} \& M.~{Iye},
  463--474

\bibitem[{{Graham} {et~al.}(2007){Graham}, {Kalas}, \& {Matthews}}]{gkm07}
{Graham}, J.~R., {Kalas}, P.~G., \& {Matthews}, B.~C. 2007, \apj, 654, 595

\bibitem[{{Hales} {et~al.}(2006){Hales}, {Gledhill}, {Barlow}, \&
  {Lowe}}]{hgb06}
{Hales}, A.~S., {Gledhill}, T.~M., {Barlow}, M.~J., \& {Lowe}, K.~T.~E. 2006,
  \mnras, 365, 1348

\bibitem[{{Harrington} \& {Kuhn}(2008)}]{hk08}
{Harrington}, D.~M., \& {Kuhn}, J.~R. 2008, \pasp, 120, 89

\bibitem[{{Henyey} \& {Greenstein}(1941)}]{hg41}
{Henyey}, L.~G., \& {Greenstein}, J.~L. 1941, \apj, 93, 70

\bibitem[{{Hinkley} {et~al.}(2008){Hinkley}, {Oppenheimer}, {Brenner}, {Parry},
  {Sivaramakrishnan}, {Soummer}, \& {King}}]{hob08}
{Hinkley}, S., {Oppenheimer}, B.~R., {Brenner}, D., {Parry}, I.~R.,
  {Sivaramakrishnan}, A., {Soummer}, R., \& {King}, D. 2008, in Society of
  Photo-Optical Instrumentation Engineers (SPIE) Conference Series, Vol. 7015,
  Society of Photo-Optical Instrumentation Engineers (SPIE) Conference Series

\bibitem[{{Hinkley} {et~al.}(2007){Hinkley}, {Oppenheimer}, {Soummer},
  {Sivaramakrishnan}, {Roberts}, {Kuhn}, {Makidon}, {Perrin}, {Lloyd},
  {Kratter}, \& {Brenner}}]{hos07}
{Hinkley}, S., {Oppenheimer}, B.~R., {Soummer}, R., {Sivaramakrishnan}, A.,
  {Roberts}, Jr., L.~C., {Kuhn}, J., {Makidon}, R.~B., {Perrin}, M.~D.,
  {Lloyd}, J.~P., {Kratter}, K., \& {Brenner}, D. 2007, \apj, 654, 633

\bibitem[{{Hong}(1985)}]{h85}
{Hong}, S.~S. 1985, \aap, 146, 67

\bibitem[{{Jura}(1991)}]{j91}
{Jura}, M. 1991, \apjl, 383, L79+

\bibitem[{{Kalas} {et~al.}(2008){Kalas}, {Graham}, {Chiang}, {Fitzgerald},
  {Clampin}, {Kite}, {Stapelfeldt}, {Marois}, \& {Krist}}]{kgc08}
{Kalas}, P., {Graham}, J.~R., {Chiang}, E., {Fitzgerald}, M.~P., {Clampin}, M.,
  {Kite}, E.~S., {Stapelfeldt}, K., {Marois}, C., \& {Krist}, J. 2008, Science,
  322, 1345

\bibitem[{{Kalas} {et~al.}(2005){Kalas}, {Graham}, \& {Clampin}}]{kgc05}
{Kalas}, P., {Graham}, J.~R., \& {Clampin}, M. 2005, \nat, 435, 1067

\bibitem[{{Keller}(2002)}]{k02}
{Keller}, C.~U. 2002, in Astrophysical Spectropolarimetry, ed.
  J.~{Trujillo-Bueno}, F.~{Moreno-Insertis}, \& F.~{S{\'a}nchez}, 303--354

\bibitem[{{Kelsall} {et~al.}(1998){Kelsall}, {Weiland}, {Franz}, {Reach},
  {Arendt}, {Dwek}, {Freudenreich}, {Hauser}, {Moseley}, {Odegard},
  {Silverberg}, \& {Wright}}]{kwf98}
{Kelsall}, T., {Weiland}, J.~L., {Franz}, B.~A., {Reach}, W.~T., {Arendt},
  R.~G., {Dwek}, E., {Freudenreich}, H.~T., {Hauser}, M.~G., {Moseley}, S.~H.,
  {Odegard}, N.~P., {Silverberg}, R.~F., \& {Wright}, E.~L. 1998, \apj, 508, 44

\bibitem[{{Kenyon} {et~al.}(1999){Kenyon}, {Wood}, {Whitney}, \&
  {Wolff}}]{kww99}
{Kenyon}, S.~J., {Wood}, K., {Whitney}, B.~A., \& {Wolff}, M.~J. 1999, \apjl,
  524, L119

\bibitem[{{Kim} {et~al.}(1994){Kim}, {Martin}, \& {Hendry}}]{kmh94}
{Kim}, S.-H., {Martin}, P.~G., \& {Hendry}, P.~D. 1994, \apj, 422, 164

\bibitem[{{Kimura} {et~al.}(2006){Kimura}, {Kolokolova}, \& {Mann}}]{kkm06}
{Kimura}, H., {Kolokolova}, L., \& {Mann}, I. 2006, \aap, 449, 1243

\bibitem[{{Koerner} {et~al.}(1998){Koerner}, {Ressler}, {Werner}, \&
  {Backman}}]{krw98}
{Koerner}, D.~W., {Ressler}, M.~E., {Werner}, M.~W., \& {Backman}, D.~E. 1998,
  \apjl, 503, L83+

\bibitem[{{K{\"o}hler} {et~al.}(2008){K{\"o}hler}, {Mann}, \& {Li}}]{kml08}
{K{\"o}hler}, M., {Mann}, I., \& {Li}, A. 2008, \apjl, 686, L95

\bibitem[{{Krist} {et~al.}(2000){Krist}, {Stapelfeldt}, {M{\'e}nard},
  {Padgett}, \& {Burrows}}]{ksm00}
{Krist}, J.~E., {Stapelfeldt}, K.~R., {M{\'e}nard}, F., {Padgett}, D.~L., \&
  {Burrows}, C.~J. 2000, \apj, 538, 793

\bibitem[{{Kuhn} {et~al.}(2001){Kuhn}, {Potter}, \& {Parise}}]{kpp01}
{Kuhn}, J.~R., {Potter}, D., \& {Parise}, B. 2001, \apjl, 553, L189

\bibitem[{{Lafreni{\`e}re} {et~al.}(2007){Lafreni{\`e}re}, {Doyon}, {Marois},
  {Nadeau}, {Oppenheimer}, {Roche}, {Rigaut}, {Graham}, {Jayawardhana},
  {Johnstone}, {Kalas}, {Macintosh}, \& {Racine}}]{ldm07}
{Lafreni{\`e}re}, D., {Doyon}, R., {Marois}, C., {Nadeau}, D., {Oppenheimer},
  B.~R., {Roche}, P.~F., {Rigaut}, F., {Graham}, J.~R., {Jayawardhana}, R.,
  {Johnstone}, D., {Kalas}, P.~G., {Macintosh}, B., \& {Racine}, R. 2007, \apj,
  670, 1367

\bibitem[{{Lucas} {et~al.}(2009){Lucas}, {Hough}, {Bailey}, {Tamura}, {Hirst},
  \& {Harrison}}]{lhb09}
{Lucas}, P.~W., {Hough}, J.~H., {Bailey}, J.~A., {Tamura}, M., {Hirst}, E., \&
  {Harrison}, D. 2009, \mnras, 393, 229

\bibitem[{{Lyot}(1939)}]{l39}
{Lyot}, B. 1939, \mnras, 99, 580

\bibitem[{{Macintosh} {et~al.}(2006){Macintosh}, {Graham}, {Palmer}, {Doyon},
  {Gavel}, {Larkin}, {Oppenheimer}, {Saddlemyer}, {Wallace}, {Bauman}, {Evans},
  {Erikson}, {Morzinksi}, {Phillion}, {Poyneer}, {Sivaramakrishnan}, {Soummer},
  {Thibault}, \& {Veran}}]{mgp06}
{Macintosh}, B., {Graham}, J., {Palmer}, D., {Doyon}, R., {Gavel}, D.,
  {Larkin}, J., {Oppenheimer}, B., {Saddlemyer}, L., {Wallace}, J.~K.,
  {Bauman}, B., {Evans}, J., {Erikson}, D., {Morzinksi}, K., {Phillion}, D.,
  {Poyneer}, L., {Sivaramakrishnan}, A., {Soummer}, R., {Thibault}, S., \&
  {Veran}, J.-P. 2006, in Advances in Adaptive Optics II. Edited by Brent L.
  Ellerbroek and Domenico Bonaccini Calia. Proceedings of the SPIE, Volume
  6272, pp. 627209 (2006).

\bibitem[{{Macintosh} {et~al.}(2008){Macintosh}, {Graham}, {Palmer}, {Doyon},
  {Dunn}, {Gavel}, {Larkin}, {Oppenheimer}, {Saddlemyer}, {Sivaramakrishnan},
  {Wallace}, {Bauman}, {Erickson}, {Marois}, {Poyneer}, \& {Soummer}}]{mgp08}
{Macintosh}, B.~A., {Graham}, J.~R., {Palmer}, D.~W., {Doyon}, R., {Dunn}, J.,
  {Gavel}, D.~T., {Larkin}, J., {Oppenheimer}, B., {Saddlemyer}, L.,
  {Sivaramakrishnan}, A., {Wallace}, J.~K., {Bauman}, B., {Erickson}, D.~A.,
  {Marois}, C., {Poyneer}, L.~A., \& {Soummer}, R. 2008, in Presented at the
  Society of Photo-Optical Instrumentation Engineers (SPIE) Conference, Vol.
  7015, Society of Photo-Optical Instrumentation Engineers (SPIE) Conference
  Series

\bibitem[{{Malbet}(1996)}]{m96}
{Malbet}, F. 1996, \aaps, 115, 161

\bibitem[{{Marois} {et~al.}(2003){Marois}, {Doyon}, {Nadeau}, {Racine}, \&
  {Walker}}]{mdn03}
{Marois}, C., {Doyon}, R., {Nadeau}, D., {Racine}, R., \& {Walker}, G.~A.~H.
  2003, in EAS Publications Series, ed. C.~{Aime} \& R.~{Soummer}, 233--243

\bibitem[{{Marois} {et~al.}(2006){Marois}, {Lafreni{\`e}re}, {Doyon},
  {Macintosh}, \& {Nadeau}}]{mld06}
{Marois}, C., {Lafreni{\`e}re}, D., {Doyon}, R., {Macintosh}, B., \& {Nadeau},
  D. 2006, \apj, 641, 556

\bibitem[{{Marois} {et~al.}(2008){Marois}, {Macintosh}, {Barman}, {Zuckerman},
  {Song}, {Patience}, {Lafreni{\`e}re}, \& {Doyon}}]{mmb08}
{Marois}, C., {Macintosh}, B., {Barman}, T., {Zuckerman}, B., {Song}, I.,
  {Patience}, J., {Lafreni{\`e}re}, D., \& {Doyon}, R. 2008, Science, 322, 1348

\bibitem[{{Metchev} \& {Hillenbrand}(2009)}]{mh09}
{Metchev}, S.~A., \& {Hillenbrand}, L.~A. 2009, \apjs, 181, 62

\bibitem[{{Nielsen} {et~al.}(2008){Nielsen}, {Close}, {Biller}, {Masciadri}, \&
  {Lenzen}}]{ncb08}
{Nielsen}, E.~L., {Close}, L.~M., {Biller}, B.~A., {Masciadri}, E., \&
  {Lenzen}, R. 2008, \apj, 674, 466

\bibitem[{{Oppenheimer} {et~al.}(2008){Oppenheimer}, {Brenner}, {Hinkley},
  {Zimmerman}, {Sivaramakrishnan}, {Soummer}, {Kuhn}, {Graham}, {Perrin},
  {Lloyd}, {Roberts}, \& {Harrington}}]{obh08}
{Oppenheimer}, B.~R., {Brenner}, D., {Hinkley}, S., {Zimmerman}, N.,
  {Sivaramakrishnan}, A., {Soummer}, R., {Kuhn}, J., {Graham}, J.~R., {Perrin},
  M., {Lloyd}, J.~P., {Roberts}, Jr., L.~C., \& {Harrington}, D.~M. 2008, \apj,
  679, 1574

\bibitem[{{Oppenheimer} {et~al.}(2004){Oppenheimer}, {Digby}, {Newburgh},
  {Brenner}, {Shara}, {Mey}, {Mandeville}, {Makidon}, {Sivaramakrishnan},
  {Soummer}, {Graham}, {Kalas}, {Perrin}, {Roberts}, {Kuhn}, {Whitman}, \&
  {Lloyd}}]{odn04}
{Oppenheimer}, B.~R., {Digby}, A.~P., {Newburgh}, L., {Brenner}, D., {Shara},
  M., {Mey}, J., {Mandeville}, C., {Makidon}, R.~B., {Sivaramakrishnan}, A.,
  {Soummer}, R., {Graham}, J.~R., {Kalas}, P., {Perrin}, M.~D., {Roberts},
  L.~C., {Kuhn}, J.~R., {Whitman}, K., \& {Lloyd}, J.~P. 2004, in Advancements
  in Adaptive Optics. Edited by Domenico B. Calia, Brent L. Ellerbroek, and
  Roberto Ragazzoni. Proceedings of the SPIE, Volume 5490, pp. 433-442 (2004).,
  ed. D.~{Bonaccini Calia}, B.~L. {Ellerbroek}, \& R.~{Ragazzoni}, 433--442

\bibitem[{{Oppenheimer} {et~al.}(2003){Oppenheimer}, {Sivaramakrishnan}, \&
  {Makidon}}]{osm03}
{Oppenheimer}, B.~R., {Sivaramakrishnan}, A., \& {Makidon}, R.~B. 2003,
  {Imaging Exoplanets: The Role of Small Telescopes} (The Future of Small
  Telescopes In The New Millennium.~Volume III - Science in the Shadow of
  Giants), 155--+

\bibitem[{{Perrin} {et~al.}(2004){Perrin}, {Graham}, {Kalas}, {Lloyd}, {Max},
  {Gavel}, {Pennington}, \& {Gates}}]{pgk04}
{Perrin}, M.~D., {Graham}, J.~R., {Kalas}, P., {Lloyd}, J.~P., {Max}, C.~E.,
  {Gavel}, D.~T., {Pennington}, D.~M., \& {Gates}, E.~L. 2004, Science, 303,
  1345

\bibitem[{{Perrin} {et~al.}(2008){Perrin}, {Graham}, \& {Lloyd}}]{pgl08}
{Perrin}, M.~D., {Graham}, J.~R., \& {Lloyd}, J.~P. 2008, \pasp, 120, 555

\bibitem[{{Perrin} {et~al.}(2002){Perrin}, {Graham}, {Trumpis}, {Kuhn},
  {Whitman}, {Coulter}, {Lloyd}, \& {Roberts}}]{p02}
{Perrin}, M.~D., {Graham}, J.~R., {Trumpis}, M., {Kuhn}, J.~R., {Whitman}, K.,
  {Coulter}, R., {Lloyd}, J.~P., \& {Roberts}, L.~C. 2002, in 2002 AMOS
  Technical Conference, P. W. Kervin, J. L. Africano; eds.

\bibitem[{{Pinte} {et~al.}(2007){Pinte}, {Fouchet}, {M{\'e}nard}, {Gonzalez},
  \& {Duch{\^e}ne}}]{pfm07}
{Pinte}, C., {Fouchet}, L., {M{\'e}nard}, F., {Gonzalez}, J.-F., \&
  {Duch{\^e}ne}, G. 2007, \aap, 469, 963

\bibitem[{{Potter}(2003)}]{p03}
{Potter}, D.~E. 2003, PhD thesis, AA(UNIVERSITY OF HAWAI'I)

\bibitem[{{Potter} {et~al.}(2000){Potter}, {Close}, {Roddier}, {Roddier},
  {Graves}, \& {Northcott}}]{pcr00}
{Potter}, D.~E., {Close}, L.~M., {Roddier}, F., {Roddier}, C., {Graves}, J.~E.,
  \& {Northcott}, M. 2000, \apj, 540, 422

\bibitem[{{Racine} {et~al.}(1999){Racine}, {Walker}, {Nadeau}, {Doyon}, \&
  {Marois}}]{rwn99}
{Racine}, R., {Walker}, G.~A.~H., {Nadeau}, D., {Doyon}, R., \& {Marois}, C.
  1999, \pasp, 111, 587

\bibitem[{{Roberts} \& {Neyman}(2002)}]{rn02}
{Roberts}, L.~C., \& {Neyman}, C.~R. 2002, \pasp, 114, 1260

\bibitem[{{Schneider} {et~al.}(1999){Schneider}, {Smith}, {Becklin}, {Koerner},
  {Meier}, {Hines}, {Lowrance}, {Terrile}, {Thompson}, \& {Rieke}}]{ssb99}
{Schneider}, G., {Smith}, B.~A., {Becklin}, E.~E., {Koerner}, D.~W., {Meier},
  R., {Hines}, D.~C., {Lowrance}, P.~J., {Terrile}, R.~J., {Thompson}, R.~I.,
  \& {Rieke}, M. 1999, \apjl, 513, L127

\bibitem[{{Schneider} {et~al.}(2009){Schneider}, {Weinberger}, {Becklin},
  {Debes}, \& {Smith}}]{swb09}
{Schneider}, G., {Weinberger}, A.~J., {Becklin}, E.~E., {Debes}, J.~H., \&
  {Smith}, B.~A. 2009, \aj, 137, 53

\bibitem[{{Sivaramakrishnan} {et~al.}(2001){Sivaramakrishnan}, {Koresko},
  {Makidon}, {Berkefeld}, \& {Kuchner}}]{skm01}
{Sivaramakrishnan}, A., {Koresko}, C.~D., {Makidon}, R.~B., {Berkefeld}, T., \&
  {Kuchner}, M.~J. 2001, \apj, 552, 397

\bibitem[{{Sivaramakrishnan} {et~al.}(2002){Sivaramakrishnan}, {Lloyd},
  {Hodge}, \& {Macintosh}}]{slh02}
{Sivaramakrishnan}, A., {Lloyd}, J.~P., {Hodge}, P.~E., \& {Macintosh}, B.~A.
  2002, \apjl, 581, L59

\bibitem[{{Sivaramakrishnan} {et~al.}(2007){Sivaramakrishnan}, {Oppenheimer},
  {Hinkley}, {Brenner}, {Soummer}, {Mey}, {Lloyd}, {Perrin}, {Graham},
  {Makidon}, {Roberts}, \& {Kuhn}}]{soh07}
{Sivaramakrishnan}, A., {Oppenheimer}, B.~R., {Hinkley}, S., {Brenner}, D.,
  {Soummer}, R., {Mey}, J.~L., {Lloyd}, J.~P., {Perrin}, M.~D., {Graham},
  J.~R., {Makidon}, R.~B., {Roberts}, L.~C., \& {Kuhn}, J.~R. 2007, Comptes
  Rendus Physique, 8, 355

\bibitem[{{Soummer} {et~al.}(2007){Soummer}, {Ferrari}, {Aime}, \&
  {Jolissaint}}]{sfa07}
{Soummer}, R., {Ferrari}, A., {Aime}, C., \& {Jolissaint}, L. 2007, \apj, 669,
  642

\bibitem[{{Soummer} {et~al.}(2006){Soummer}, {Oppenheimer}, {Hinkley},
  {Sivaramakrishnan}, {Makidon}, {Digby}, {Brenner}, {Kuhn}, {Perrin},
  {Roberts}, \& {Kratter}}]{soh06}
{Soummer}, R., {Oppenheimer}, B.~R., {Hinkley}, S., {Sivaramakrishnan}, A.,
  {Makidon}, R.~B., {Digby}, A.~P., {Brenner}, D., {Kuhn}, J.~R., {Perrin},
  M.~D., {Roberts}, L.~C., \& {Kratter}, K. 2006, in EAS Publications Series,
  ed. C.~{Aime} \& M.~{Carbillet}

\bibitem[{{Sparks} \& {Ford}(2002)}]{sf02}
{Sparks}, W.~B., \& {Ford}, H.~C. 2002, \apj, 578, 543

\bibitem[{{Stapelfeldt} {et~al.}(2004){Stapelfeldt}, {Holmes}, {Chen}, {Rieke},
  {Su}, {Hines}, {Werner}, {Beichman}, {Jura}, {Padgett}, {Stansberry},
  {Bendo}, {Cadien}, {Marengo}, {Thompson}, {Velusamy}, {Backus}, {Blaylock},
  {Egami}, {Engelbracht}, {Frayer}, {Gordon}, {Keene}, {Latter}, {Megeath},
  {Misselt}, {Morrison}, {Muzerolle}, {Noriega-Crespo}, {Van Cleve}, \&
  {Young}}]{shc04}
{Stapelfeldt}, K.~R., {Holmes}, E.~K., {Chen}, C., {Rieke}, G.~H., {Su},
  K.~Y.~L., {Hines}, D.~C., {Werner}, M.~W., {Beichman}, C.~A., {Jura}, M.,
  {Padgett}, D.~L., {Stansberry}, J.~A., {Bendo}, G., {Cadien}, J., {Marengo},
  M., {Thompson}, T., {Velusamy}, T., {Backus}, C., {Blaylock}, M., {Egami},
  E., {Engelbracht}, C.~W., {Frayer}, D.~T., {Gordon}, K.~D., {Keene}, J.,
  {Latter}, W.~B., {Megeath}, T., {Misselt}, K., {Morrison}, J.~E.,
  {Muzerolle}, J., {Noriega-Crespo}, A., {Van Cleve}, J., \& {Young}, E.~T.
  2004, \apjs, 154, 458

\bibitem[{{Stokes}(1852)}]{s1852}
{Stokes}, G.~G. 1852, Trans. Camb. Phil. Soc., 3, 399

\bibitem[{{Tinbergen}(1996)}]{t96}
{Tinbergen}, J. 1996, {Astronomical Polarimetry}, ed. J.~{Tinbergen}

\bibitem[{{van Leeuwen}(2007)}]{v07}
{van Leeuwen}, F. 2007, \aap, 474, 653

\bibitem[{{Wahhaj} {et~al.}(2007){Wahhaj}, {Koerner}, \& {Sargent}}]{wks07}
{Wahhaj}, Z., {Koerner}, D.~W., \& {Sargent}, A.~I. 2007, \apj, 661, 368

\bibitem[{{Wood} {et~al.}(2002){Wood}, {Lada}, {Bjorkman}, {Kenyon}, {Whitney},
  \& {Wolff}}]{wlb02}
{Wood}, K., {Lada}, C.~J., {Bjorkman}, J.~E., {Kenyon}, S.~J., {Whitney}, B.,
  \& {Wolff}, M.~J. 2002, \apj, 567, 1183

\end{thebibliography}
\bibliographystyle{/Users/shinkley/Library/texmf/tex/latex/apj} 

%To do: 
%cite ZIMPOL paper, Beuzit Messenger,  GISLER spie paper. 
%cite Debes.
%Interpretation: how does your g value relate to Debes-like studies. 
%spell check 
%exposure times on figure 5 plot. 

\end{document}